\documentclass[3p,times,procedia]{elsarticle}
\usepackage{nupha_ecrc}
\usepackage{slashed}
\usepackage{stackengine}

\newcommand\onecirc{\stackMath\mathbin{\stackinset{c}{0ex}{c}{0ex}{1}{\bigcirc}}}
\newcommand\twocirc{\stackMath\mathbin{\stackinset{c}{0ex}{c}{0ex}{2}{\bigcirc}}}

\volume{00}

\firstpage{1}

\journalname{Nuclear Physics A}

\runauth{}

\jid{nupha}

\jnltitlelogo{Nuclear Physics A}

\usepackage{amssymb}
\usepackage{amsmath}

\usepackage[figuresright]{rotating}

\begin{document}

\begin{frontmatter}



\dochead{XXVIIIth International Conference on Ultrarelativistic Nucleus-Nucleus Collisions\\ (Quark Matter 2019)}

\title{Chirality transfer \& chiral turbulence in gauge theories}


\author[label1,label2]{Mark Mace}
\address[label1]{Department of Physics, University of Jyv\"{a}skyl\"{a}, P.O. Box 35, 40014, Jyv\"{a}skyl\"{a}, Finland}
\address[label2]{Helsinki Institute of Physics, University of Helsinki, P.O. Box 64, 00014, Helsinki, Finland}
\author[label3]{Niklas Mueller}
\address[label3]{Physics Department, Brookhaven National Laboratory, Bldg. 510A, Upton, NY 11973, USA}
\author[label4]{S\"{o}ren Schlichting}
\address[label4]{Fakult\"{a}t f\"{u}r Physik, Universit\"{a}t Bielefeld, D-33615 Bielefeld, Germany}
\author[label5]{Sayantan Sharma}
\address[label5]{The Institute of Mathematical Sciences, HBNI, Chennai 600113, India}

\runauth{M.Mace, N.Mueller, S.Schlichting and S.Sharma}

\begin{abstract}
Chirality transfer between fermions and gauge fields plays a crucial role for understanding the dynamics of anomalous transport phenomena such as the Chiral Magnetic Effect. In this proceeding we present a first principles study of these processes based on classical-statistical real-time lattice simulations of strongly coupled QED $(e^2N_f=64)$. Our simulations demonstrate that a chirality imbalance in the fermion sector triggers chiral plasma instabilities in the gauge field sector, which ultimately lead to the generation of long range helical magnetic fields via an self-similar turbulent cascade of the magnetic helicity.
\end{abstract}


\end{frontmatter}


\section{Introduction \& Motivation}
Chiral transport phenomena, such as the Chiral Magnetic Effect (CME) \cite{Fukushima:2008xe} have recently inspired a lot of theoretical and experimental attention with expected manifestations across the most diverse range of energy scales, from cosmology \cite{Brandenburg:2017rcb} and heavy-ion collisions \cite{Skokov:2016yrj} all the way to Dirac/Weyl semi-metals in condensed matter physics \cite{Li:2014bha}. Despite the fact that a macroscopic description of chiral transport phenomena can be achieved within the framework of anomalous hydrodynamics \cite{Son:2009tf}, there remains one crucial difference in comparison to ordinary transport phenomena, such as e.g. energy-momentum transport or vector charge diffusion. While ordinary transport phenomena are associated with the transport of conserved quantities, such as the energy-momentum tensor $T^{\mu\nu}$ or vector current $j^{\mu}_{v}$, chiral transport phenomena are linked to the dynamics of the chiral current $j^{\mu}_{5}= (n_5,\mathbf{j}_{5})$ which is not conserved due to quantum anomalies. Chiral transport phenomena are thus intrinsically non-equilibrium effects, and a robust theoretical description inevitably requires an understanding of the dynamics of axial charge changing processes in the system.

Generally, the non-conservation of the chiral current is expressed by the $U(1)_{A}$ anomaly relation, which for a gauge theories with $N_{f}$ flavors of fundamental fermions takes the form
\begin{eqnarray}
\partial_{\mu}j^{\mu}_{5}(t,\mathbf{x})= 2  \sum_{f} \bar{\psi}_{f}(t,\mathbf{x}) m_{f} i\gamma_{5} \psi_{f}(t,\mathbf{x}) - 2 N_{f} \partial_{\mu} K^{\mu}(t,\mathbf{x})\;.
\end{eqnarray}
Here $\partial_{\mu} j^{\mu}_{5}$ is the covariant divergence of the chiral current of fermions, and $\partial_{\mu}K^{\mu}(t,\mathbf{x}) = \frac{g^2}{16\pi^2} \text{tr}[ F_{\mu\nu}(t,\mathbf{x}) \tilde{F}^{\mu\nu}(t,\mathbf{x})]$ represents the covariant divergence of the chiral current of the gauge fields. Denoting the net-chirality of the gauge fields as $N_{h}(t)= \int d^3\mathbf{x}~K^{0}(t,\mathbf{x})$, which for $\text{SU}(N_c)$ gauge theories (QCD) represents the Chern-Simons number and respectively for $\text{U}(1)$ gauge theories (QED) the magnetic helicity, one finds that in the chiral limit $m_{f} \to 0$ the net-chirality of the system  $N_{5}(t)+2N_f N_{h}(t)$ is conserved, indicating that the chiral charge can be continuously exchanged between fermions and gauge fields. 

Chirality transfer from gauge fields to fermions can be most easily achieved via the application of parallel external electric and magnetic fields (QED) \cite{Li:2014bha}, but is also possible due to space-time dependent fluctuations of (chromo-) electro-magnetic fields (QED \& QCD) \cite{Mace:2016svc,Figueroa:2019jsi} and in non-abelian gauge theories via sphaleron transitions between different topological sectors (QCD) \cite{Moore:2010jd,Mace:2016svc}. Conversely, the transfer of chirality from fermions to gauge fields may occur e.g. due to a bias of the sphaleron transition rate  (QCD) \cite{McLerran:1990de} or due to the presence of the chiral plasma instabilities (QED \& QCD) \cite{Akamatsu:2013pjd,Hirono:2015rla} in chirality imbalanced plasmas. Besides the fact that a detailed understanding of these processes is necessary in order to establish a valid macroscopic description of chiral transport in heavy-ion collisions, it is also interesting to point out that chiral plasma instabilities in particular may have interesting consequences for the dynamics of $U(1)$ gauge fields in Astrophysics and Cosmology \cite{Yamamoto:2015gzz,Rogachevskii:2017uyc,Brandenburg:2017rcb,Schober:2017cdw}.

\begin{figure}[t!]
\centering
\includegraphics[width=\textwidth]{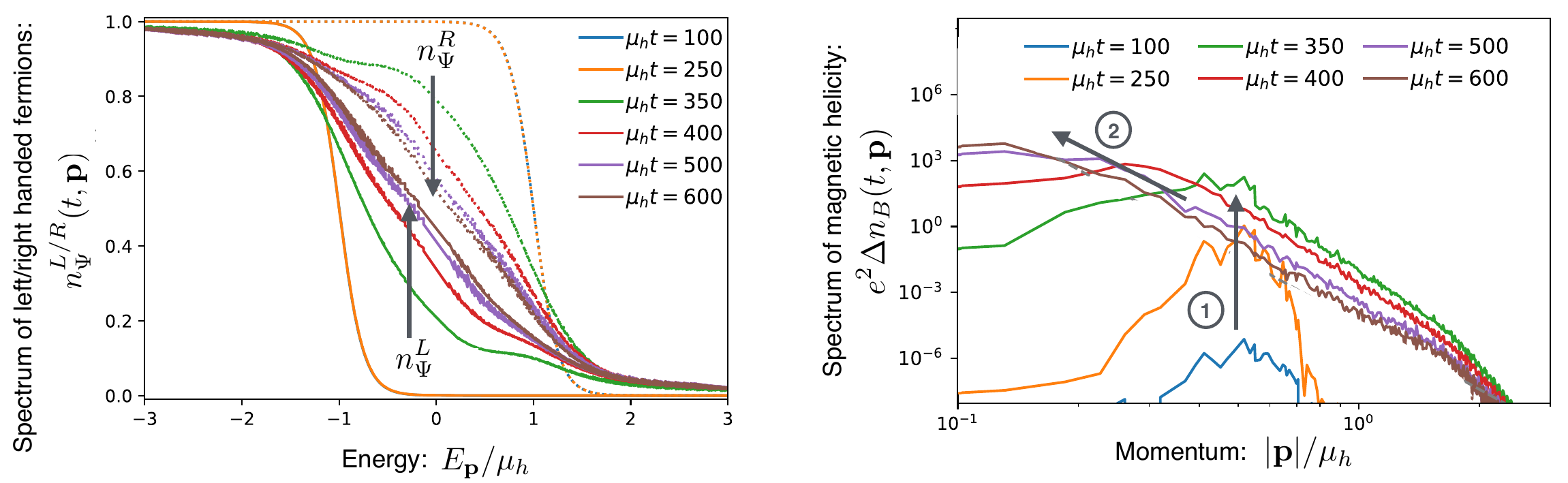}
\caption{Evolution of the occupation numbers of left/right handed fermion spectra $n_{\Psi}^{L/R}(t,\mathbf{p})$ (left) and the magnetic helicity $\Delta n_{B}(t,\mathbf{p})=n_{B}^{R}(t,\mathbf{p}) - n_{B}^{L}(t,\mathbf{p})$ (right). Starting from a chirality imbalance in the fermion sector, the chiral plasma instability leads to an exponential growth of the magnetic helicity in a narrow range of momenta. Subsequently, the magnetic helicity is transferred to low momentum modes via a self-similar inverse cascade.}
\label{fig:ChiralitySpectra}
\end{figure}

\section{Chirality transfer \& chiral turbulence in strongly coupled QED plasmas}
We report recent progress in understanding of the dynamics of chirality transfer processes via first principles simulations of strongly coupled QED plasmas \cite{Mace:2019cqo}. Starting from a chirality imbalance in the fermion sector, which is characterized by a helicity chemical potential $\mu_{h}$ much larger than the temperature $T$, we perform classical-statistical real-time lattice simulations of the underlying quantum field dynamics by numerically solving the coupled set of Maxwell's and Dirac equations
\begin{eqnarray}
\partial_{\mu}F^{\mu\nu}(t,\mathbf{x}) = eN_f \left\langle \bar{\hat{\Psi}}(t,\mathbf{x}) \gamma^{\mu} \hat{\Psi}(t,\mathbf{x}) \right \rangle\;, \qquad (i\slashed{D}-m_{f})\hat{\Psi}(t,\mathbf{x})=0\;.
\end{eqnarray}
By discretizing the theory on a $N_s \times N_s \times N_s$ spatial lattice with lattice spacing $a_s$ in each direction, the operator equation for $\hat{\Psi}(t,\mathbf{x})$ becomes finite dimensional and can be solved numerically based on a mode function expansion \cite{Mace:2016shq}, albeit the cost of the computation scales as $N_s^6$. Details of the simulation procedure are provided in \cite{Mace:2019cqo} and \cite{Mace:2016shq}, and we only note that we employ a compact lattice discretization of the $U(1)$ gauge fields along with $\mathcal{O}(a_s^3)$ improved Wilson-Dirac fermions. Simulations are performed for strongly coupled plasmas with $e^2N_f=64$ to properly resolve all dynamical scales on the available size lattices up to $N_s=48$ and $a_s=1$. Even though it is straightforward to extend our simulations to finite quark mass, we focus on the dynamics close to the chiral limit with $m_f \ll \mu_{h}$, where the net-chirality of fermions and gauge fields $N_{5}(t)+2N_f N_{h}(t)$ is effectively conserved over the course of the evolution.

\begin{figure}[t!]
\centering
\includegraphics[width=\textwidth]{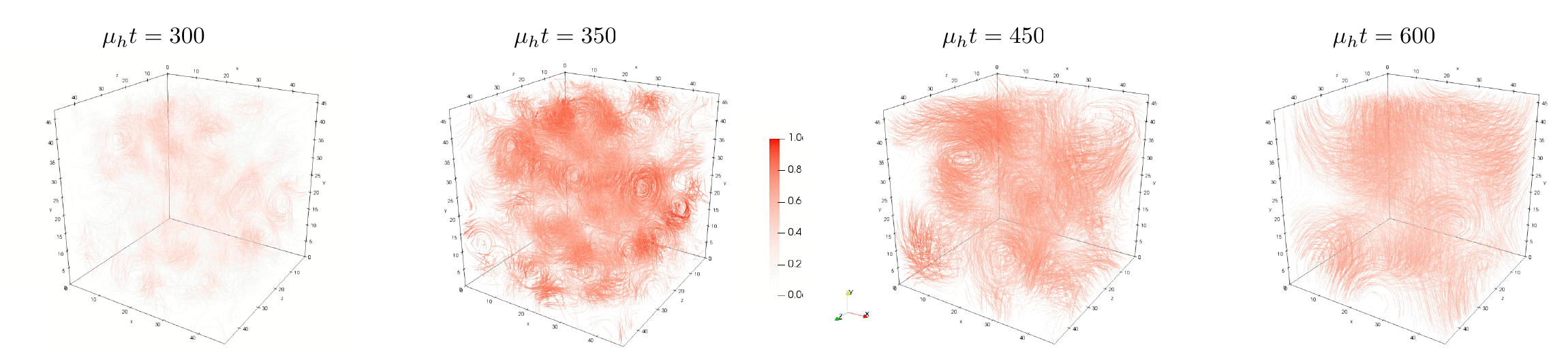}
\caption{Evolution of magnetic fields $\mathbf{B}(t,\mathbf{x})$ generated from the initial chirality imbalance in the fermion sector. Different panels show stream-tracing plots of the magnetic field lines at different stages of the evolution, during the linear instability regime $\mu_{h} t=300$, close to the saturation of unstable growth $\mu_{h} t=350$ and during the self-similar turbulent regime $\mu_h t=450,600$. Generation of large scale magnetic fields is achieved via the self-similar inverse cascade of magnetic helicity. }
\label{fig:BFields}
\end{figure}

Starting from a chirality imbalance in the fermion sector, we find that chirality transfer and the generation of large scale helical magnetic fields proceeds in two steps, which are compactly summarized in Fig. \ref{fig:ChiralitySpectra} where we present results for the evolution of the magnetic helicity spectra
\begin{eqnarray}
\Delta n_{B}(t,\mathbf{p})=n_{B}^{R}(t,\mathbf{p}) - n_{B}^{L}(t,\mathbf{p})\;, \qquad n_{B}^{L/R}(t,\mathbf{p})=\frac{|\mathbf{B}^{L/R}(t,\mathbf{p})|^{2}}{|\mathbf{p}|}\;, \qquad \mathbf{B}^{L/R}(t,\mathbf{p})= \frac{|\mathbf{p}|\pm i \mathbf{p} \times }{2|\mathbf{p}|} \mathbf{B}(t,\mathbf{p})
\end{eqnarray}
in the right panel, along with the evolution of the occupation numbers of left/right handed fermion spectra $n_{\Psi}^{L/R}(t,\mathbf{p})$ in the left panel (see \cite{Mace:2019cqo} for details). Due to the chirality imbalance in the fermion sector, right-handed magnetic fields are initially subject to the chiral plasma instability, and exhibit and exponential growth within a narrow range of momenta around $p_{\rm inst} \simeq 0.5 \mu_{h}$ as indicated by the $\onecirc$ arrow in Fig. \ref{fig:ChiralitySpectra}. Exponential growth terminates around $\mu_{h}t^{*} \approx 350$ when the occupancies of the magnetic fields become non-perturbatively large $e^2 n_{B}^{R}(t,\mathbf{p}) \sim 1$ and the spectrum of magnetic fields spreads out towards the ultra-violet and the infrared. Simultaneously, a significant fraction of the chiral charge is transferred from fermions to gauge fields, as can be seen from the left panel of Fig.~\ref{fig:ChiralitySpectra} in terms in of the narrowing of the gap in the fermi surface between left- and right handed fermions. Subsequently, the plasma enters a turbulent regime, where magnetic helicity is transferred from high momentum modes with $p \simeq p_{\rm inst}$ to low momentum modes with $p \ll p_{\rm inst}$ via an inverse cascade, indicated by the $\twocirc$ arrow in Fig. \ref{fig:ChiralitySpectra}. We find that during the turbulent phase, i.e. for $\mu_{h} t \gtrsim \mu_{h}t^{*} \approx 350$, the spectrum of the magnetic helicity follows a self-similar scaling behavior of the form
\begin{eqnarray}
\label{eq:selfsimilar}
e^2\Delta n_{B}(t,\mathbf{p})=\left[\mu_{h} (t-t^{*})\right]^{\alpha}f_{s}\left([\mu_{h} (t-t^{*})]^{\beta} |\mathbf{p}|\right)
\end{eqnarray}
with stationary spectrum $f_{s}$ and scaling exponents $\alpha=1.14 \pm 0.5$ and $\beta=0.37 \pm 0.13$ extracted from a statistical scaling analysis. Since the inverse cascade in Eq.~(\ref{eq:selfsimilar}) is associated with the transport of the magnetic helicity 
\begin{eqnarray}
N_{h}(t)= V \int \frac{d^3\mathbf{p}}{(2\pi)^3} e^2 \Delta n_{B}(t,\mathbf{p})\;, \qquad V=N_s^3 a_S^3\;,
\end{eqnarray}
from high momentum to low momentum modes, it eventually results in the generation of long range helical magnetic fields, as can be seen from Fig.~\ref{fig:BFields} where we present snapshots of the magnetic field configurations at different times. Over the course of the chiral plasma instability, strong magnetic fields with short correlation length are generated from the chirality imbalance in the fermion sector up to times $\mu_{h} t \lesssim 350$ when exponential growth terminates. Subsequently, for $\mu_{h} t \gtrsim 350$ the inverse cascade leads to a spatial growth of magnetic field domains, eventually generating the long range magnetic fields seen in the right most panel of Fig.~\ref{fig:BFields}. 

\section{Conclusions \& Outlook}
We performed real-time lattice simulations of strongly coupled QED plasmas to investigate chirality transfer processes in gauge theories. We find that the chiral plasma instability provides an efficient mechanism to transfer chiral charge from fermions to gauge fields, and eventually leads to the generation of large scale helical magnetic fields via self-similar (inverse) cascade of magnetic helicity. While our microscopic simulations qualitatively confirm the picture established in macroscopic studies based on anomalous magneto-hydrodynamics \cite{Hirono:2015rla,Schober:2017cdw}, there are also some important differences concerning in particular the evolution at late times. While previous studies of chirally imbalanced plasmas with $\mu_{h} \ll T$ suggest that exponential growth terminates due to a depletion of the chirality imbalance of fermions, our simulations suggest that in the presence of a large chirality imbalance $\mu_{h} \gg T$ saturation of the instability occurs when magnetic fields become non-perturbatively large $e^2 n_{B}^{R}(t,\mathbf{p}) \sim 1$, resulting in strong inhomogeneities of the plasma and enhanced generation of large scale magnetic fields. Evidently a detailed comparison between the different scenarios would be desirable, as clarification of these aspects will be important for future progress.

So far the results reported in this proceeding have been obtained for a $U(1)$ gauge theory (QED) close to the chiral limit; extending these studies to include dissipative effects due to finite fermion masses is straightforward and will be explored in future studies. Similarly, it will be important to perform analogous studies for non-Abelian $SU(N_c)$ gauge theories (QCD), to clarify the role of chirality transfer processes in the theoretical description of anomalous transport phenomena in heavy-ion collisions.

\textit{Acknowledgements:}
We gratefully acknowledge support for this work by the European Research Council under grant ERC-2015-CoG-681707 (M.M.), by the U.S. Department of Energy, Office of Science, Office of Nuclear Physics, under contract No. DE- SC0012704 (N.M.), by the Deutsche Forschungsgemeinschaft (DFG, German Research Foundation)  through Project 404640738 (N.M.) and CRC-TR 211 “Strong-interaction matter under extreme conditions” Project number 315477589 (S.S.), and by the Department of Science and Technology, Govt. of India through a Ramanujan fellowship and from the Institute of Mathematical Sciences (Sa.S.). This research used resources of the National Energy Research Scientific Computing Center (NERSC), a U.S. Department of Energy Office of Science User Facility operated under Contract No. DE-AC02-05CH11231.





\bibliographystyle{elsarticle-num}


\end{document}